\documentclass{evn2004}
\usepackage{txfonts}
\usepackage{graphicx}
\usepackage{natbib}
\begin{document}
   \title{Spectral Properties of the Core and the VLBI-Jets of Cygnus\,A}

   \author{U. Bach, T.P. Krichbaum, E. Middelberg, M. Kadler, W. Alef, A. Witzel and J.A. Zensus
          }
   \authorrunning{U. Bach et al.}
   \institute{Max-Planck-Institut f\"ur Radioastronomie, Auf dem H\"ugel 69,
   53121 Bonn, Germany} 

   \abstract{We present a detailed VLBI study of the spectral properties of the inner core region of the radio galaxy Cygnus\,A at 5\,GHz, 15\,GHz, 22\,GHz, 43\,GHz and 86\,GHz. Our observations include an epoch using phase-referencing
at 15\,GHz and 22\,GHz and the first successful VLBI observations of Cygnus A at 86 GHz. We find a pronounced two-sided jet structure, with a steep spectrum along the jet and an inverted spectrum towards the counter-jet. The inverted spectrum and the frequency-dependent jet-to-counter-jet ratio suggest that the inner counter-jet is covered by a circum-nuclear absorber as it is proposed by the unified scheme.
   }
   \maketitle
%
%

\paragraph{\large\textbf{Introduction}}
Cygnus\,A is the closest ($z=0.057$) strong FR\,II radio galaxy and therefore a
key object for detailed studies of AGN. Its kiloparsec-scale structure in the
radio bands is dominated by two prominent radio lobes which contain bright
hot-spots. The radio core lies in the centre of an elliptical galaxy
(\citealt{1974MNRAS.166..305H}) and is powering the two thin jets and the radio
lobes (e.g., \citealt{1984ApJ...285L..35P}). Very Long Baseline Interferometry
(VLBI) images from 1.6\,GHz to 43\,GHz obtained during the last 20 years
(\citealt{1991AJ....102.1691C,1993A&A...275..375K,1994AJ....108...64C,1998A&A...329..873K,2002evlb.conf..155B,Bach2004})
revealed a pronounced two-sided core-jet structure also on parsec scales.
According to the unified scheme, narrow line radio galaxies, like Cyg\,A, should
contain an obscuring torus around the central engine that blocks the emission from
the broad line region (BLR) (e.g. \citealt{1995PASP..107..803U}).  Evidence for a
hidden BLR in Cyg\,A comes from UV spectroscopy (\citealt{1994Natur.371..313A})
and optical spectro-polarimetry (\citealt{1997ApJ...482L..37O}). Their results are
supported by the detection of H\,I absorption  near the core and on the counter-jet
side (\citealt{1995ApJ...449L.131C}). The idea of a ring or disk-like free-free
absorber surrounding the nucleus is further supported from our multi-frequency
VLBI studies, which show an inverted spectrum of the counter-jet and a
frequency dependent jet-to-counter jet flux density ratio (Krichbaum et al. 1998;
Bach et al. 2002). 

In this study and with new data, we obtain further constrains for the
circum-nuclear absorber. Here we show new spectral index maps of the innermost
portion of the jet of Cyg\,A, obtained at frequencies from 5\,GHz to 86\,GHz.\\

\paragraph{\large\textbf{Observations and Data Reduction}}\label{cyg_sec:obs}

We carried out six multi-frequency VLBI epochs of Cyg\,A between 1996 and 2003, including a phase-referencing observations at 15\,GHz and 22\,GHz. We obtained the first VLBI image of Cyg\,A at 86\,GHz, using the technique of fast frequency
switching (\citealt{2002evlb.conf...61M}). A detailed observing log is given in
Tab.~\ref{cyg_tab:obslog}.

\begin{table}[htbp]
\caption[Cygnus\,A observing log.]{Observation log.}
\label{cyg_tab:obslog}
\scriptsize
\centering
\begin{tabular}{lcclccl}
\hline
Epoch & 
\multicolumn{1}{c}{$\nu$} & 
\multicolumn{1}{c}{$S_{\rm tot}$} & 
\multicolumn{1}{c}{Beam, P.A.} &
\multicolumn{1}{c}{$S_{\rm peak}$} & 
\multicolumn{1}{c}{$\sigma$} &
\multicolumn{1}{c}{Pol.}\\
 & 
\multicolumn{1}{c}{[GHz]} & 
\multicolumn{1}{c}{[Jy]} & 
\multicolumn{1}{c}{[mas\,$\times$\,mas], [$^\circ$]} &
\multicolumn{1}{c}{$[\frac{Jy}{beam}]$} & 
\multicolumn{1}{c}{$[\frac{mJy}{beam}]$} &~\vspace{1pt}\\
\hline
1996.73\,$^{\rm a}$ & 15.4 & 1.71 & $0.30 \times 0.61$, $-18.3$ & 0.40 & 0.36 & dual\\
1996.73\,$^{\rm a}$ & 22.2 & 1.48 & $0.24 \times 0.47$, $-16.1$ & 0.36 & 0.82 & dual\\
1996.73\,$^{\rm a}$ & 43.2 & 1.01 & $0.23 \times 0.27$,\hspace{8pt} $3.6$ & 0.40 & 1.90 & dual\\
2002.03\,$^{\rm a}$ &  4.9 & 0.89 & $0.92 \times 1.54$, $-23.8$ & 0.12 & 0.17 & dual\\
2002.03\,$^{\rm a}$ & 15.4 & 1.51 & $0.31 \times 0.56$, $-21.3$ & 0.28 & 0.12 & dual\\
2002.51\,$^{\rm a}$ &  4.9 & 0.91 & $0.87 \times 1.56$, $-23.0$ & 0.15 & 0.16 & dual\\
2002.51\,$^{\rm a}$ & 15.4 & 1.50 & $0.46 \times 0.67$, $-14.4$ & 0.39 & 0.17 & dual\\
2003.04\,$^{*}$ & 15.4 & 1.27 & $0.46 \times 0.73$,\hspace{4pt} $ -5.1$ & 0.32 & 0.26 & dual\\
2003.04\,$^{*}$ & 22.2 & 1.27 & $0.31 \times 0.51$,\hspace{4.5pt} $10.4$ & 0.34 & 0.45 & dual\\
2003.24\,$^{\rm a}$ &  4.9 & 0.99 & $1.10 \times 1.72$, $-20.1$ & 0.21 & 0.14 & dual\\
2003.24\,$^{\rm a}$ & 15.4 & 1.38 & $0.25 \times 0.52$, $-23.4$ & 0.23 & 0.12 & dual\\
2003.27         & 14.4 & 1.52 & $0.45 \times 0.68$,\hspace{8pt} $0.6$ & 0.35 & 0.26 &  LCP\\
2003.27         & 43.1 & 0.75 & $0.16 \times 0.26$, $-11.4$ & 0.23 & 0.64 &  LCP\\
2003.27         & 86.2 & 0.41 & $0.32 \times 0.36$,\hspace{4.5pt} $ 88.4$ & 0.33 & 1.89 &  LCP\\
\hline		
\end{tabular}
\flushleft
\scriptsize
Note: The array used was the VLBA, unless indicated by a footnote. Epochs in
bold face denote own data. a: VLBA+Eb. *:
phase-referencing. Listed are the observing
epoch, frequency $\nu$, total flux density $S_{\rm tot}(\nu)$, beam size, beam
position angle, peak flux density $S_{\rm peak}$, $\sigma$ the rms noise of the
map and polarization mode.
\end{table}

The data were reduced in the standard manner using NRAO's Astronomical Image
Processing System ({\sc Aips}). The imaging of the source employing phase and
amplitude self-calibration was done using the CLEAN and SELFCAL procedures in {\sc
Difmap}. The self-calibration was done  in steps of several phase-calibrations
followed by careful amplitude calibration. A more detailed description of the data
reduction can be found in \cite{BachPhD}.

\paragraph{\large\textbf{Results and Discussion}}\label{cyg_sec:results}

\begin{figure}[htbp]
\centering
\includegraphics[bb=0 0 714 766,angle=0,width=0.93\columnwidth,clip] {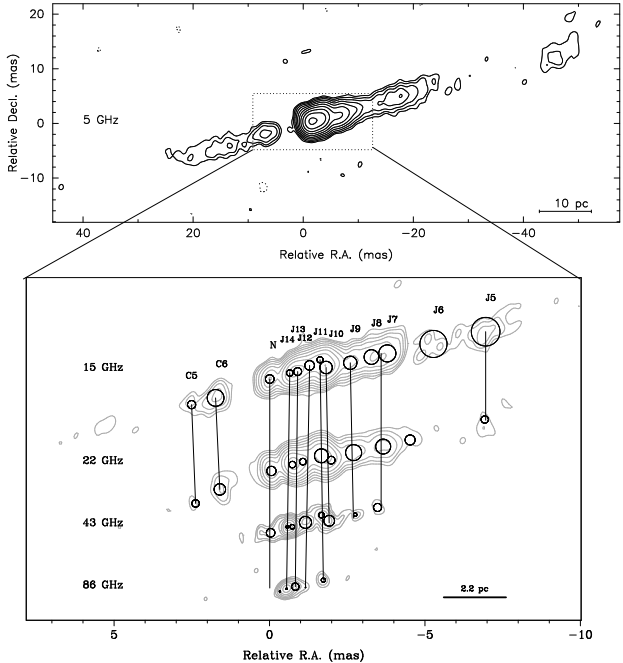}
   \caption[VLBI images of Cygnus\,A]{VLBI images of Cyg\,A at 5\,GHz (2003.24), 15\,GHz and 22\,GHz (2003.04, phase-ref.), 43\,GHz (2003.27) and 86\,GHz (2003.27). The beam size, peak flux density and rms noise are given
   Tab.~\ref{cyg_tab:obslog}. The lowest contours start at 3$\sigma $
   and increase by steps of 2. The bottom panel also shows the overlaid modelfit components and their identification between different frequencies.}\label{cyg_fig:maps}
\vspace{-0.5cm}
\end{figure}

A collection of images obtained at different frequencies during 2003 is shown in
Fig.~\ref{cyg_fig:maps}. At 5\,GHz the jet and the counter-jet extend up to $\sim
50$\,mas from the core (1\,mas $\approx 1.1$\,pc). The width of the gap of
emission located $\sim1- 4$\,mas east of the intensity maximum decreases with
frequency, indicative of strong opacity effects on the counter-jet side.

The cross-identification was done using individual modelfit components along the
jets. The identification was done using their relative separation from each other,
their flux density and size. An upper limit of 0.2\,mas for the shift of the
brightest component between 15\,GHz and 22\,GHz could be derived from the
phase-referencing observation (2003.04) and was used to constrain the
identification. Component N was used as a reference point, and can be naively
interpreted as being the nucleus. Our analysis, however, shows that it is more likely
the first counter-jet component while the true centre of activity might be located
between N and J14 (Fig. 1 \& \citealt{BachPhD}).

\begin{figure}[htbp]
\centering
\includegraphics[bb=48 52 705 522,angle=0,width=0.95\columnwidth,clip] {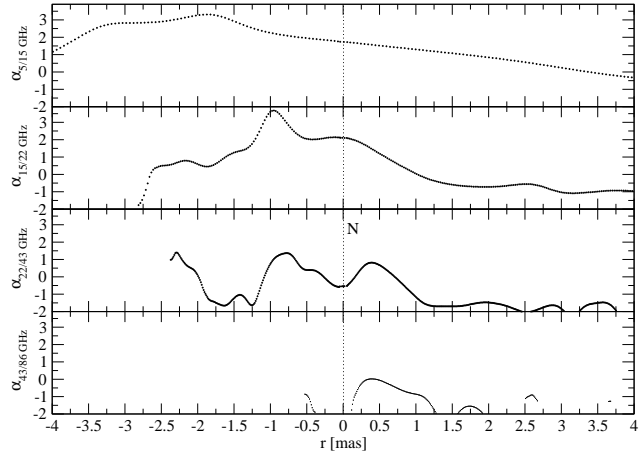}
 \caption[Spectral index profiles between 5, 15, 22, 43, and 86\,GHz (2003.2)]
 {Spectral index profiles between frequency pairs at (5, 15, 22, 43, and 86)\,GHz. The profiles represent cuts through spectral index maps along the
 ridge-line of the jet. The dotted line marks
 the position of component N.}\label{cyg_fig:spix_profile} 
\end{figure}

Spectral index profiles of the inner region around the core are presented in
Fig.~\ref{cyg_fig:spix_profile} and show clearly the different behaviour of the
spectral index at different frequency pairs. Most of the jet emission has a steep
spectrum, whereas the counter-jet spectrum is flatter. The spectral properties in
the core region, ($-2 \leq r \leq 1$)\,mas, are much more complex, showing a
highly inverted spectrum between 5\,GHz and 15\,GHz and also highly inverted
regions at the higher frequency pairs, but always at different locations.

Synchrotron self-absorption can produce spectral indices of up to 2.5, but between
$(-3\leq r\leq0)$\,mas the spectral index between 5\,GHz and 15\,GHz exceeds this
maximum significantly. The  most likely explanation is that in this region the
lower frequencies become affected by free-free absorption as observed, e.g.,
prominently in the lower-luminosity system NGC\,1052 (e.g., Kadler et al. 2004).
Recent simulations of radiative transfer models for obscuring tori in active
galaxies which were applied to Cyg\,A (\citealt{2003A&A...404....1V}) show that
the spectral energy distribution (SED) is best fitted by an inclined ($\sim
50^\circ$) torus of 10\,pc to 30\,pc, which is in good agreement with our results.

We estimate the absorption using fits of free-free absorbed synchrotron emission
to the jet spectra. At the positions where we think free-free absorption is most
relevant the resulting optical depth at 5\,GHz is on average $4.3\pm1.0$. Assuming
a typical temperature of 10$^4$\,K and a path length of about 5\,pc for the
absorber, this correspond to an absorbing column density of  $\sim 7 \times
10^{23}$\,cm, which is within the uncertainties of the measurements in good
agreement with the column density inferred from X-ray absorption of $4 \times
10^{23}$\,cm$^{-2}$ (\citealt{2002ApJ...564..176Y}).

To summarize, the emission gap between the jet and the counter-jet at 5\,GHz seems
to be the imprint of a circum-nuclear absorber which might cover also a large
fraction of the counter-jet up to 20\,mas and becomes optically thin further out,
where the counter-jet shines through. 
\begin{acknowledgements}
We thank the group of the VLBA 2cm Survey for providing their data. This work
made use of the VLBA, which is an instrument of the National Radio Astronomy
Observatory, a facility of the National Science Foundation, operated under
cooperative agreement by Associated Universities, Inc.\ and of the 100\,m
telescope at Effelsberg, which is operated by the Max-Planck-Institut f\"ur
Radioastronomie in Bonn. U.B.\ acknowledges partial support from the EC ICN RadioNET. M.K.\ is supported for this research through a stipend from the International Max Planck Research School (IMPRS) for Radio and Infrared Astronomy at the University of Bonn.
\end{acknowledgements}
%

\begin{thebibliography}{}
\footnotesize

\bibitem[{{Antonucci} {et~al.}(1994){Antonucci}, {Hurt}, \&
  {Kinney}}]{1994Natur.371..313A}
{Antonucci}, R., et al. 1994, \nat, 371, 313

\bibitem[{{Bach}(2004)}]{BachPhD}
{Bach}, U. 2004, PhD thesis, Rheinische Friedrich-Wilhelms-Universit\"at Bonn

\bibitem[{{Bach} {et~al.}(2004){Bach}, {Kadler}, {Krichbaum}, {Middelberg},
  {Alef}, {Witzel}, \& {Zensus}}]{Bach2004}
{Bach}, U., et al. 2004, in Future Directions
  in High Resolution Astronomy: The 10th Anniversary of the VLBA, ed.
  J.~{Romney} \& M.J.~{Reid} (ASP), in press, astro-ph/0309403

\bibitem[{{Bach} {et~al.}(2002){Bach}, {Krichbaum}, {Alef}, {Witzel}, \&
  {Zensus}}]{2002evlb.conf..155B}
{Bach}, U., et al.
  2002, in Proceedings of the 6th EVN Symposium, ed. E.~{Ros}, et al. (Bonn, Germany: MPIfR), 155


\bibitem[{{Carilli} {et~al.}(1994){Carilli}, {Bartel}, \&
  {Diamond}}]{1994AJ....108...64C}
{Carilli}, C.~L.,  et al. 1994, \aj, 108, 64

\bibitem[{{Carilli} {et~al.}(1991){Carilli}, {Bartel}, \&
  {Linfield}}]{1991AJ....102.1691C}
{Carilli}, C.~L.,  et al. 1991, \aj, 102, 1691

\bibitem[{{Conway} \& {Blanco}(1995)}]{1995ApJ...449L.131C}
{Conway}, J.~E. \& {Blanco}, P.~R. 1995, \apjl, 449, L131


\bibitem[{{Hargrave} \& {Ryle}(1974)}]{1974MNRAS.166..305H}
{Hargrave}, P.~J. \& {Ryle}, M. 1974, \mnras, 166, 305

\bibitem[{{Kadler} {et al.} (2004)}]{KadlerNGC1052} {Kadler}, M., et al. 2004 A\&A, in press, astro-ph/0407283

\bibitem[{{Krichbaum} {et~al.}(1998){Krichbaum}, {Alef}, {Witzel}, {Zensus},
  {Booth}, {Greve}, \& {Rogers}}]{1998A&A...329..873K}
{Krichbaum}, T.~P., et al. 1998, \aap, 329, 873

\bibitem[{{Krichbaum} {et~al.}(1993){Krichbaum}, {Witzel}, {Graham}, {Standke},
  {Schwartz}, {Lochner}, {Schalinski}, {Greve}, {Steppe}, {Brunswig}, {Butin},
  {Hein}, {Navarro}, {Penalver}, {Grewing}, {Booth}, {Colomer}, \&
  {Ronnang}}]{1993A&A...275..375K}
{Krichbaum}, T.~P.,  et al. 1993, \aap, 275,
  375

\bibitem[{{Middelberg} {et~al.}(2002){Middelberg}, {Roy}, {Walker}, {Falcke},
  \& {Krichbaum}}]{2002evlb.conf...61M}
{Middelberg}, E., et al. 2002, in Proceedings of the 6th EVN Symposium, ed. E.~{Ros}, et al. (Bonn, Germany: MPIfR), 61

\bibitem[{{Ogle} {et~al.}(1997){Ogle}, {Cohen}, {Miller}, {Tran}, {Fosbury}, \&
  {Goodrich}}]{1997ApJ...482L..37O}
{Ogle}, P.~M., et al. 1997, \apjl, 482, L37

\bibitem[{{Perley} {et~al.}(1984){Perley}, {Dreher}, \&
  {Cowan}}]{1984ApJ...285L..35P}
{Perley}, R.~A.,  et al. 1984, \apjl, 285, L35


\bibitem[{{Urry} \& {Padovani}(1995)}]{1995PASP..107..803U}
{Urry}, C.~M. \& {Padovani}, P. 1995, \pasp, 107, 803

\bibitem[{{van Bemmel} \& {Dullemond}(2003)}]{2003A&A...404....1V}
{van Bemmel}, I.~M. \& {Dullemond}, C.~P. 2003, \aap, 404, 1

\bibitem[{{Young} {et~al.}(2002){Young}, {Wilson}, {Terashima}, {Arnaud}, \&
  {Smith}}]{2002ApJ...564..176Y}
{Young}, A.~J., et al. 2002, \apj, 564, 176

\end{thebibliography}

\end{document}